\documentclass[pdftex,preprint,twocolumn]{revtex4}
\usepackage[english]{babel}
\usepackage{amsmath}
\usepackage{times}
\usepackage{hyperref}
\usepackage[]{graphicx}

\newcommand{\So}{\rho_{600}}
\newcommand{\Eo}{E_{0}}
\newcommand{\Ro}{R_{0}}
\newcommand{\Rms}{R_{\text{m.s.}}}
\newcommand{\dgr}{^{\text{o}}}

\begin{document}

\title{A shape of charged particle lateral distribution in individual
  EAS events with energy above $10^{19}$~eV arriving from different
  celestial regions}

\author{A. V. Sabourov}\email[]{tema@ikfia.ysn.ru}
\author{M. I. Pravdin}\email[]{m.i.pravdin@ikfia.ysn.ru}
\author{S. P. Knurenko}\email[]{s.p.knurenko@ikfia.ysn.ru}

\affiliation{Yu. G. Shafer Institute of Cosmophysical Research and
  Aeronomy, 31 Lenin Ave., 677980 Yakutsk, Russia}

\begin{abstract}
  A shape of lateral distribution for charged particles in
  events with energy above $10^{19}$~eV is considered. Two methods
  were used for individual LDF parametrization. In the first approach,
  the index of power was determined for generalized Greisen-Linsley
  approximation. In second, mean square radius of the shower was
  determined for approximation proposed by Lagutin et al. Comparison
  of resulted parameters is presented for individual events arrived
  from different celestial regions~--- Galactic planes and the region
  with increased flux of particles with $E_{0} \ge 10^{19}$~eV
  (according to Yakutsk array): $1.7\text{h} - 3.7\text{h}$ right
  ascension ; $45\dgr - 60\dgr$ declination.
\end{abstract}

\maketitle

\section{Introduction}

The knowledge of the lateral distribution function (LDF) of charged
particles from extensive air shower (EAS) is vital for experiments in
the field of ultra-high energy cosmic ray (UHECR) studying. It is LDF
that defines main shower parameters such as $\So$~ (charged particle
density at the distance $600$~m from the core) and thus~--- primary
energy.

In this paper we consider parameters of individual LDFs resulted from
revision of high energy events registered at the Yakutsk EAS
array. The aim of this work is to trace possible correlation between
parameters of individual showers and their arrival directions on the
sky, especially for Galactic planes and for the region with
significantly increased UHECR flux, detected by Yakutsk
group~\cite{bib:Ivanov}.

\section{Estimation of lateral distribution parameters for individual
  showers}

For the analysis we selected showers with $\Eo \ge 10^{19}$~eV, with
zenith angles $\theta < 60\dgr$ and with core lying well within the
boundaries of the array, to make sure that shower core is found
correctly.

At the Yakutsk EAS array, approximation proposed by Greisen
~\cite{bib:NKG} is used for primary data processing:
\begin{equation}
\rho(r) = M \cdot \left(\frac{r}{\Ro}\right)^{-1} \cdot \left(1 +
  \frac{r}{\Ro}\right)^{\left<b\right> + 1} \text{,}
\label{eq:NKG}
\end{equation}
where $\Ro$ is Moiere radius and slope parameter $\left<b\right> = -
1.38 - 2.16 \cdot \cos{\theta} - 0.15 \cdot \lg{\So}$.

In the work by Glushkov et al~\cite{bib:Glushkov}, an updated
approximation was proposed, that demonstrated better description of
experimental points at large distances from the core ($r > 1000$~m):
\begin{equation}
\begin{split}
\rho(r) = & M \cdot \left(\frac{r}{\Ro}\right)^{-1.3} \cdot \left(1 +
  \frac{r}{\Ro}\right)^{\left<b\right> + 1.3} \times{}\\
  & \times \left(1 + \frac{r}{2000}\right)^{-3.5} \text{,}
\label{eq:Glushkov}
\end{split}
\end{equation}
where $\left<b\right> = 2.6 \cdot (1 - \cos{\theta}) - 3.242$.

In equations (\ref{eq:NKG}) and (\ref{eq:Glushkov}) the slope parameter
$\left<b\right>$ is derived from average LDF. While it describes most
of showers quite well, it certainly fails doing so in dozen number of
events. During revision we performed $\chi^{2}$-fitting of functions
(\ref{eq:NKG}) and (\ref{eq:Glushkov}) normalized to $\So$ on
experimental data for each selected shower with free parameters $\So$
and $b$.

\begin{figure}
\centering
\includegraphics[width=0.45\textwidth, clip]{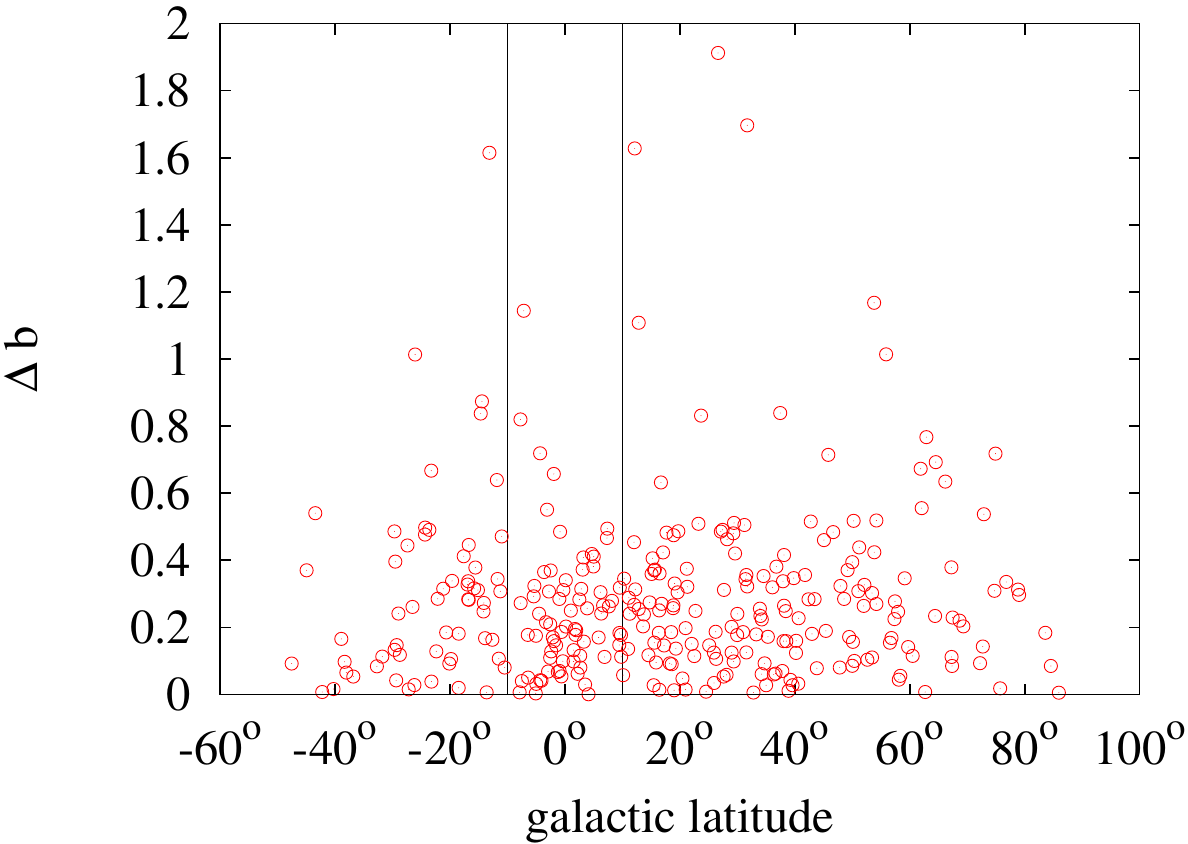}
\caption{$\Delta b$ vs Galactic latitude. Marked stripe $\pm 10\dgr$
  is Galaxy plane.}
\label{slope_galaxy}
\end{figure}

The value $\Delta b = \left|\left<b\right> - b\right|$ could give a
hint of possible astrophysical aspect of the slope parameter in
functions (\ref{eq:NKG}) and (\ref{eq:Glushkov}). As seen on
Fig.\ref{slope_galaxy}, comparison to Galactic coordinates showed no
correlation between $\Delta b$ and Galaxy plane.

To trace possible dependency on astrophysical coordinates, we selected
a ``stripe'' of $15\dgr$ width along declination and divided it into
``chunks'' of $2$h each along right ascension. Such a stripe was
selected to exclude zenith-angular dependency. Averaged $\Delta b$
values in each chunk are presented in table~\ref{tab:slope} in
comparison to averaged value in the rest chunks of the stripe.

\begin{figure}
\centering
\includegraphics[width=0.45\textwidth, clip]{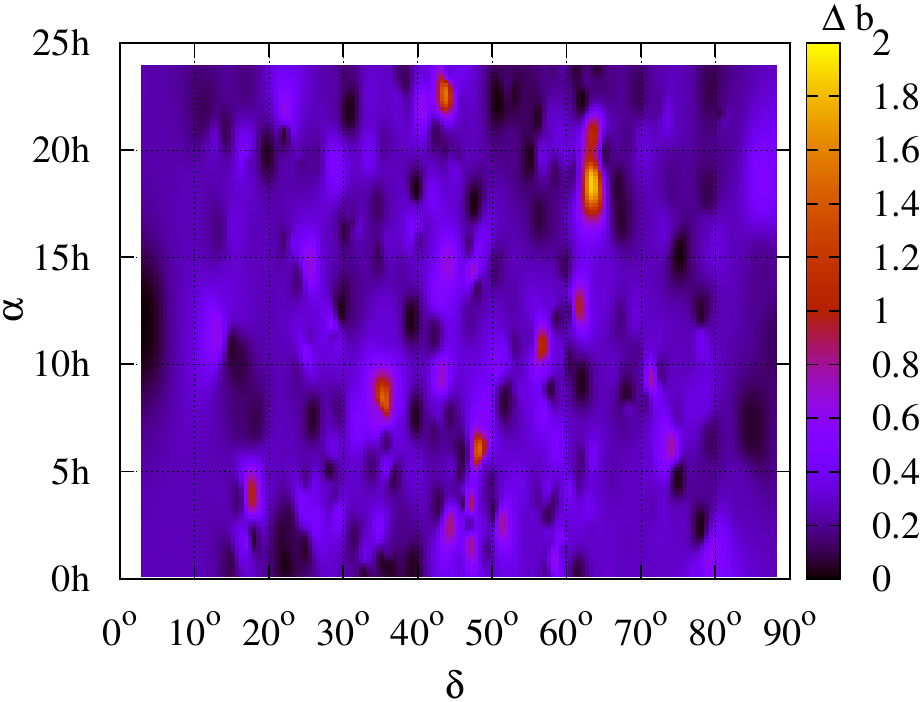}
\caption{$\Delta b$ vs astrophysical coordinates.}
\label{slope3d}
\end{figure}

\begin{table*}
\centering
\caption{Averaged $\Delta b$ values in the region of interest (see
  Fig.~\ref{slope3d})}
\label{tab:slope}
\renewcommand{\arraystretch}{1.3}
\begin{tabular}{|c|ccc|ccc|}
  \hline
  &\multicolumn{3}{|c|}{inward} & \multicolumn{3}{|c|}{outward} \\
  \cline{2-7}
  $\alpha$, hour & $\Delta b$ & $n$ & $\delta(\Delta b)$ & $\Delta b$
  & $n$ & $\delta(\Delta b)$ \\
  \hline
  $ 1.7 -  3.7$ & $0.356961$ & $15$ & $0.080848$ & $0.285764$ & $72$ &
  $0.031073$ \\
  \hline
  $ 3.7 -  5.7$ & $0.231064$ & $ 8$ & $0.041157$ & $0.304822$ & $79$ &
  $0.031815$ \\
  $ 5.7 -  7.7$ & $0.547938$ & $ 5$ & $0.283294$ & $0.282802$ & $82$ &
  $0.025840$ \\
  $ 7.7 -  9.7$ & $0.245541$ & $ 7$ & $0.081267$ & $0.302633$ & $80$ &
  $0.030999$ \\
  $ 9.7 - 11.7$ & $0.457900$ & $ 8$ & $0.109829$ & $0.281851$ & $79$ &
  $0.029796$ \\
  $11.7 - 13.7$ & $0.200063$ & $ 6$ & $0.045722$ & $0.305297$ & $81$ &
  $0.031049$ \\
  $13.7 - 15.7$ & $0.328212$ & $ 5$ & $0.134513$ & $0.296200$ & $82$ &
  $0.030067$ \\
  $15.7 - 17.7$ & $0.157237$ & $ 4$ & $0.085138$ & $0.304825$ & $83$ &
  $0.030185$ \\
  $17.7 - 19.7$ & $0.170663$ & $ 3$ & $0.102908$ & $0.302589$ & $84$ &
  $0.029958$ \\
  $19.7 - 21.7$ & $0.249221$ & $ 9$ & $0.037238$ & $0.303672$ & $78$ &
  $0.032253$ \\
  $21.7 - 23.7$ & $0.157450$ & $ 8$ & $0.043347$ & $0.312276$ & $79$ &
  $0.031443$ \\
\hline
\end{tabular}
\end{table*}

\subsection{Scaling approach}

\begin{figure}
\centering
\includegraphics[width=0.45\textwidth, clip]{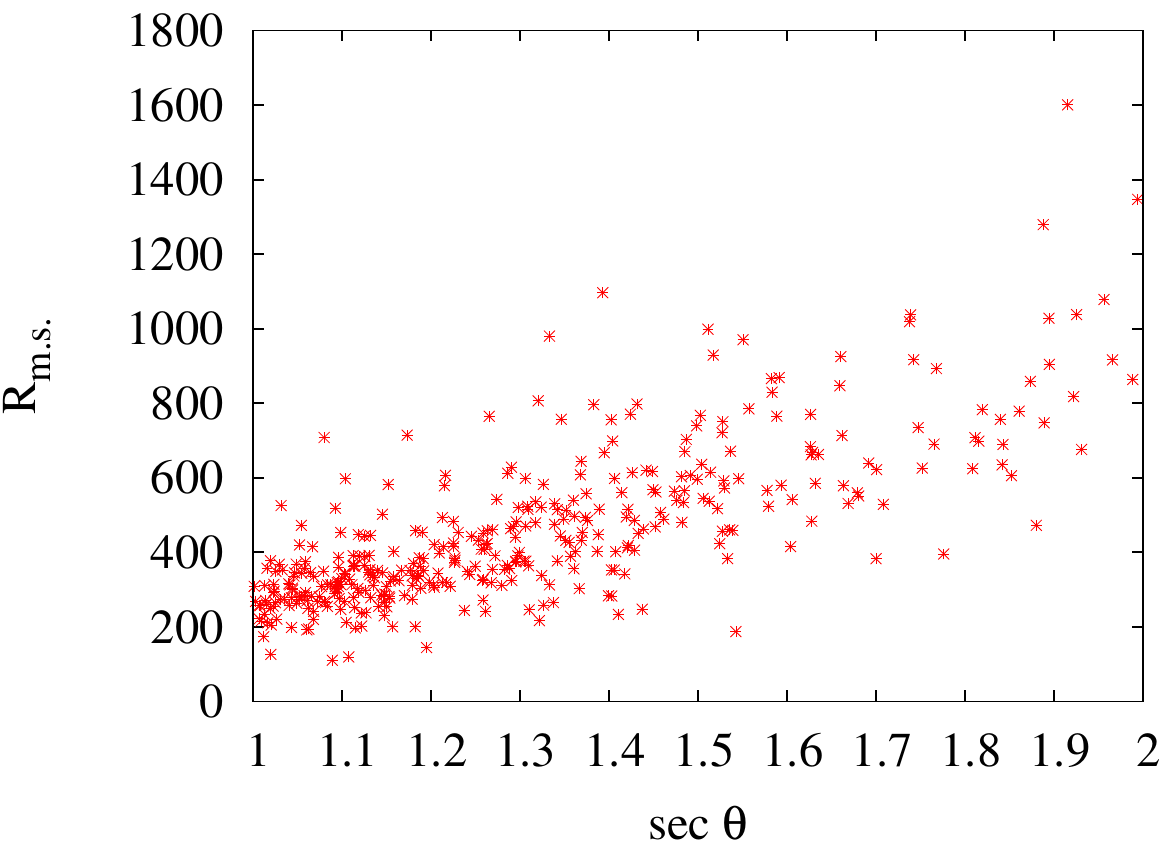}
\caption{$\Rms$ of electrons in individual showers
  compared to $\sec{\theta}$.}
\label{fig:rms_zen}
\end{figure}

A one-parametric scaling representation of charged particle lateral
distribution was proposed by Lagutin et al~\cite{bib:Lagutin}:

\begin{equation}
\begin{split}
\rho(r)  = & M \cdot \left(\frac{r}{\Rms}\right)^{-1.2} \cdot \left(1
  + \frac{r}{\Rms}\right)^{-3.33} \times \\
  & \times \left(1 + \left[\frac{r}{10 \cdot
        \Rms}\right]^{2}\right)^{-0.6} \text{,}
\label{eq:Lagutin}
\end{split}
\end{equation}
here $\Rms$ is mean square radius of electrons. This function
was obtained with respect to nuclear cascade process in the
shower~\cite{bib:Lagutin}. Since the main classification parameter for
the Yakutsk array is $\So$, we used (\ref{eq:Lagutin}) normalized to
$\So$. We calculated $\Rms$ for each shower in our selection using
$\chi^{2}$-minimization. On Fig.~\ref{fig:rms_zen} there are shown
$\Rms$ values obtained for individual events compared to zenith
angle. It is clear, that these values significantly exceed predicted
in the work~\cite{bib:Lagutin}, though one can note distinct
zenith-angular dependence.

\begin{figure}
\centering
\includegraphics[width=0.45\textwidth, clip]{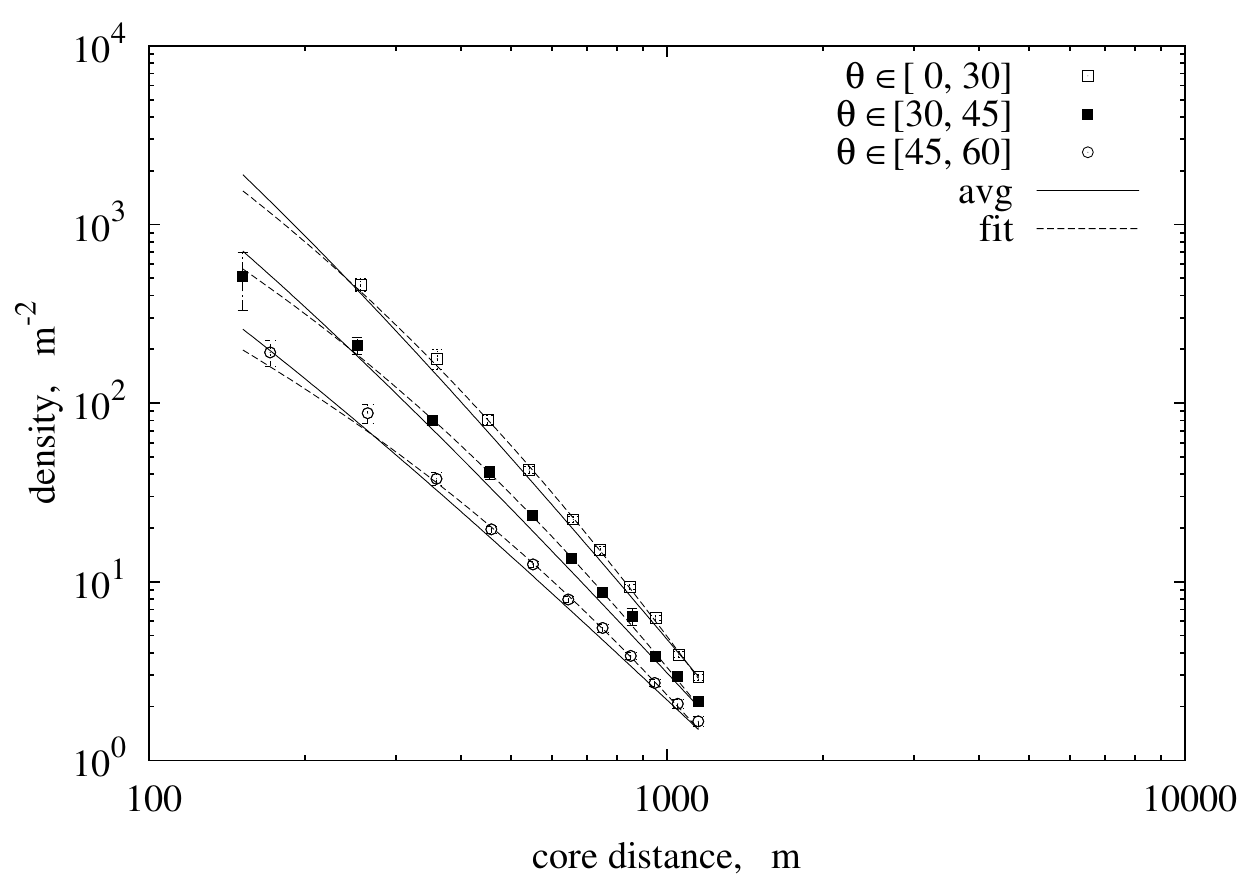}
\caption{Average scaling LDFs for three different zenith angle
  intervals.}
\label{fig:avg_ldf}
\end{figure}

We constructed average LDFs for three zenith-angular intervals: $0 -
30\dgr$, $30 - 45\dgr$ and $45 - 60\dgr$. Results can be found in
table~\ref{tab:rms} and Fig.\ref{fig:avg_ldf}. It is seen from the
table, that resulted $\Rms$ values contradict to theoretical
predictions from the work~\cite{bib:Lagutin}.

\begin{table}
\centering
\caption{Parameters for average LDF obtained for
  approximation~(\ref{eq:Lagutin})}
\label{tab:rms}
\renewcommand{\arraystretch}{1.3}
\begin{tabular}{ccccc}
\hline
$\theta$ & $\left<\So\right>$ & $\So$ & $\Rms$ & $\chi^{2}$ \\
\hline
$0 - 30\dgr$ & $26.92$ & $31.47$ & $320.42$ & $6.0594$ \\
$30 - 45\dgr$ & $14.81$ & $17.86$ & $476.45$ & $5.0713$ \\
$45 - 60\dgr$ & $8.59$  & $10.15$ & $770.07$ & $8.3343$ \\
\hline
\end{tabular}
\end{table}

Obtained $\Rms$ values did not allow us to make juxtaposition with
celestial coordinates as for functions (\ref{eq:NKG}) and
(\ref{eq:Glushkov}).

\section{Results}

Revised parameters of individual lateral distribution functions in
Greisen's ((\ref{eq:NKG}) and (\ref{eq:Glushkov})) approximation showed
no correlation neither with Galactic plane, nor with the region of
UHECR region excess. From table~\ref{tab:slope} it is seen, that
increased $\Delta b$ in the region of interest ($1.7\text{h} < \alpha
< 3.7\text{h}$) is not significant and the whole picture is spoiled by
poor statistics.

Difficulties in estimation of $\Rms$ did now allow us to use scaling
approximation~(\ref{eq:Lagutin}) in such analysis. In the
work by MSU EAS group~\cite{bib:Kalmykov} authors have faced similar
obstacles in $\Rms$ determination. It is worth mentioning, that
KASCADE-Grande group successfully used scaling formalism for
estimation of muon density in air showers~\cite{bib:KASCADE}.
Besides, scintillation detectors used at the Yakutsk array may lead to
sloping of charged particle distribution caused by registration of
atmospheric muons and electrons from muon decay. If we consider this
fact together with zenith-angular dependence of $\Rms$ more closely,
we can obtain more plausible estimation of this parameter.

\end{document}